\documentclass[sigconf, table, xcdraw, dvipsnames]{acmart}

\usepackage{multirow}
\usepackage{enumitem}
\usepackage{pifont}
\usepackage{balance}

\usepackage{twemojis}
\makeatletter
\def\@fnsymbol#1{%
  \ifcase#1\or \twemoji{cat}\or †\or ‡\or §\or ¶\or ‖\else\@ctrerr\fi}
\makeatother

\usepackage{pifont}   
\newcommand{\cmark}{\textcolor{ForestGreen}{\ding{51}}}
\newcommand{\xmark}{\textcolor{Red}{\ding{55}}}

\AtBeginDocument{%
  }

\copyrightyear{2025}
\acmYear{2025}
\setcopyright{acmlicensed}
\acmConference[RecSys '25]{Proceedings of the Nineteenth ACM Conference on Recommender Systems}{September 22--26, 2025}{Prague, Czech Republic}
\acmBooktitle{Proceedings of the Nineteenth ACM Conference on Recommender Systems (RecSys '25), September 22--26, 2025, Prague, Czech Republic}
\acmDOI{10.1145/3705328.3748164}
\acmISBN{979-8-4007-1364-4/2025/09}

\ccsdesc[500]{Information systems~Recommender systems}

\keywords{recommender systems; sequential recommendations; data splitting}

\begin{abstract}

Modern sequential recommender systems, ranging from lightweight transformer-based variants to large language models, have become increasingly prominent in academia and industry due to their strong performance in the next-item prediction task. Yet common evaluation protocols for sequential recommendations remain insufficiently developed: they often fail to reflect the corresponding recommendation task accurately, or are not aligned with real-world scenarios.

Although the widely used \textit{leave-one-out} split matches next-item prediction, it permits the overlap between training and test periods, which leads to temporal leakage and unrealistically long test horizon, limiting real-world relevance.
\textit{Global temporal splitting} addresses these issues by evaluating on distinct future periods. However, its applications to sequential recommendations remain loosely defined, particularly in terms of selecting target interactions and constructing a validation subset that provides necessary consistency between validation and test metrics.

In this paper, we demonstrate that evaluation outcomes can vary significantly across splitting strategies, influencing model rankings and practical deployment decisions. To improve reproducibility in both academic and industrial settings, we systematically compare different splitting strategies for sequential recommendations across multiple datasets and established baselines.
Our findings show that prevalent splits, such as leave-one-out, may be insufficiently aligned with more realistic evaluation strategies.

\begin{center}
  Code: \href{https://github.com/monkey0head/time-to-split}{%
    \textcolor{purple}{https://github.com/monkey0head/time-to-split}%
}
\end{center}
\end{abstract}

\begin{teaserfigure}
\vspace{-9pt}
\centering  \includegraphics[width=1.0\textwidth]{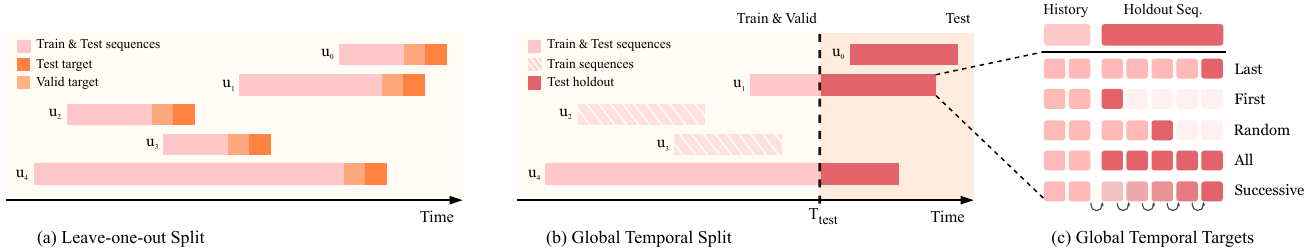}
  \caption{Data splitting and target selection strategies for sequential recommendations. (a) Leave-one-out split.
  (b) Global temporal split: all interactions after timepoint $T_{\textnormal{test}}$ are placed in the holdout set, targets for these holdout sequences are chosen according to \textnormal{(c)}. (c) Target items selection options for each holdout sequence (applicable for both test and validation sequences).}
  \label{fig:teaser}
\end{teaserfigure}

\makeatletter
\@printpermissionfalse
\makeatother

\begin{document}

\title{Time to Split: Exploring Data Splitting Strategies for Offline Evaluation of Sequential Recommenders
}

\author{Danil Gusak}
\orcid{0009-0008-1238-6533}
\affiliation{%
  \institution{AIRI, Skoltech}
  \city{Moscow}
  \country{Russian Federation}
}
\email{danil.gusak@skoltech.ru}
\authornote{Authors contributed equally to the paper}

\author{Anna Volodkevich}
\orcid{0009-0002-7958-0097}
\affiliation{%
  \institution{Sber AI Lab, Skoltech}
  \city{Moscow}
  \country{Russian Federation}
}
\email{volodkanna@yandex.ru}
\authornotemark[1]

\author{Anton Klenitskiy}
\orcid{0009-0005-8961-6921}
\affiliation{%
  \institution{Sber AI Lab}
  \city{Moscow}
  \country{Russian Federation}
}
\email{antklen@gmail.com}
\authornotemark[1]

\author{Alexey Vasilev}
\orcid{0009-0007-1415-2004}
\affiliation{%
  \institution{Sber AI Lab, HSE University}
  \city{Moscow}
  \country{Russian Federation}
}
\email{alexxl.vasilev@yandex.ru}

\author{Evgeny Frolov}
\orcid{0000-0003-3679-5311}
\affiliation{%
  \institution{AIRI, HSE University}
  \city{Moscow}
  \country{Russian Federation}
}
\email{frolov@airi.net}

\maketitle

\section{Introduction}

Sequential recommender systems (SRS) have become a prominent choice for the next-item prediction (NIP) task~\cite{zhai2024actions, ye2025fuxi}. By modeling each user’s interaction history as an ordered sequence, sequential approaches can effectively capture temporal patterns and incrementally update user representations without retraining~\cite{kang2018self, sun2019bert4rec}.

One critical component in the experimental pipeline for recommender systems (RS) is data splitting. Previous research highlighted the sensitivity of recommendation outcomes to different splitting strategies in classical recommender scenarios~\cite{meng2020exploring, sun2023take, ji2023critical}.
However, despite rapid advances in sequential architectures, evaluation protocols for SRS remain insufficiently developed: researchers commonly rely on leave-one-out (LOO) split that\textit{ violate the global timeline of user-item interactions}~\cite{hidasi2023widespread, ji2023critical}, or simply adopt global temporal split (GTS) from classical top-K recommendation task (e.g. 80/10/10 temporal split~\cite{tao2023task, li2024rechorus2}), \textit{without tailoring it to next-item prediction setting} (see Section~\ref{sec:papers_and_frame}). 
These mismatched protocols raise concerns for both the task alignment and the real-world relevance of offline evaluation results.

In this paper, we seek to close this gap and systematically examine various global temporal splitting variants \textit{specifically tailored to sequential recommendation scenarios}. We formalize and compare the prevalent LOO approach with GTS-based splits, defining the options for ground-truth target selection suited to the next-item-prediction task.
Some of these options, appeared in recent research~\cite{harte2023leveraging, frolov2024self, klenitskiy2024does}, include such NIP-oriented targets as \textit{the user's last interaction in a time-separated holdout sequence} or \textit{successive target}, where each subsequent interaction of the holdout sequence is treated as a separate target with incrementally extended input history.

Furthermore, we investigate different validation schemes for GTS, including \textit{global temporal, user-based, and last training item} splitting approaches, to analyze the trade-off between training data amount, data recency, and presence of temporal leakage in validation.
We examine the resulting subsets obtained after different splits to assess training set sizes, number of test users, durations of test and validation periods, and time-gap biases for GTS targets.

Our extensive experiments on multiple datasets and widely-used sequential models show that the choice of data splitting strategy for SRS can \textit{significantly impact evaluation metrics and model rankings}. 

In summary, \textit{our main contributions} are:
\begin{itemize}
[leftmargin=*,label=\textbullet,nosep]
    \item We explore different global temporal split variants for SRS, distinguished by choice of ground-truth targets and validation set construction, and compare them to leave-one-out split, highlighting their properties, advantages, and disadvantages;
    \item We systematically analyze metric correlations and consistency in final model rankings for different splits, identifying which strategies better align with real-world scenarios and the next-item prediction task;
    \item We evaluate different validation schemes for GTS and identify those that offer reliable model selection.
\end{itemize}

\section{Related Work}
\label{sec:relatedwork}

In this section, we first outline the evolution of SRS and then review existing studies on data splitting strategies for the offline evaluation of recommender systems.

\paragraph{Sequential Recommenders}

Early SRS research was advanced with gated RNNs~\cite{hidasi2015session, 10.1162/neco.1997.9.8.1735}, which outperformed MC models ~\cite{he2016fusing, he2017translation}. 
The emergence of Transformers~\cite{vaswani2017attention} led to further improvements, consistently surpassing prior methods~\cite{kang2018self, sun2019bert4rec}.
Subsequent development is continuing in various directions, including enriching recommendation models with side information~\cite{liu2021noninvasive, tisas, rahmani2023incorporating}, integrating contrastive learning approaches~\cite{zeng2025non, cui2024diffusion, du2022contrastive, duo}, and modifying the self-attention mechanism~\cite{light, denoise}.
Recently, generative recommendations~\cite{li2023gpt4rec, rajput2023recommender, singh2024better, volodkevich2024autoregressive} and integration of SRS with large language models~\cite{chen2024hllm, firooz2025360brew, yu2024break, xu2024slmrec} have emerged as promising directions.

While some works redesign training objectives in favor of a long-term engagement~\cite{pancha2022pinnerformer}, the next-item prediction task remains dominating in the vast majority of the works mentioned above. Given the rapid advancement and widespread use of sequential recommenders, it is critical to carefully select data splitting strategies, matching the real-world usage and recommendation task, for robust evaluation and comparison of the SRS models.

\paragraph{Data Splitting Strategies}

Reproducible evaluation remains a persistent challenge in recommender systems research, as evidenced by the findings of~\citet{Dacrema2019}, which demonstrate that only a small fraction of newly proposed algorithms consistently outperform rigorously optimized baselines. Corroborating this, \citet{hidasi2023widespread} systematically categorize how variations in experimental protocol can yield inconsistent model rankings.

Data splitting, an essential part of the evaluation protocol, has been deeply analyzed in various studies. Some works explore the impact of different evaluation settings, including data splitting, on the performance of top-N recommendation algorithms~\cite{sun2021areweevaluatingrigorouslypcore,zhao2022revisiting}. Other studies~\cite{sun2023take, ji2023critical} provide a critical analysis of data leakage in commonly used random and leave-one-out splitting strategies. Further,~\citet{sun2023take,ji2023critical} argue that a better evaluation strategy should use a global timeline to avoid data leakage and better reflect real-world scenarios. ~\citet{meng2020exploring} show that the splitting strategy can significantly affect the ranking of recommendations, making comparisons across studies difficult. It also highlights that certain splitting strategies may favor specific recommendation models. The authors use Kendall's correlation to assess the consistency of the splits. Several other studies consider different aspects of data splitting:~\citet{scheidt2021time} propose evaluating models on a sequence of consecutive time-based splits to account for how model performance changes over time;~\citet{verachtert2022we} highlight how the length of the training window influences algorithm performance; and \citet{wegmeth2023effect} investigate how randomness in data splitting affects the variability of evaluation metrics.

Unlike the works mentioned above, we focus on sequential recommendation algorithms and splitting strategies tailored for the next-item prediction task, where we observe an absence of a unified split protocol that both prevents data leakage and ensures reproducible, fair comparisons of SRS in a close-to-real usage setting.

\section{Data Splitting Strategies for Sequential Recommendations}
\label{sec:strategies}

\subsubsection{Important splitting properties considered} 
To be aligned with the real-world usage and avoid data leakage from the future, the split must preserve a \textit{global timeline}~\cite{ji2023critical}, meaning that any interaction occurring after the timepoint $T_{test}$
is excluded from training. In production, sequential recommenders are often used in online scenarios for \textit{next-item prediction} over some period (e.g., day or week) before model retraining, and \textit{all previous user history is available for inference}, including interactions after training cutoff. A splitting strategy not aligned with real-world usage can inflate offline performance and promote models that underperform in production.

Statistical tests are often used to make an offline evaluation result a reliable estimate of a model's performance~\citep{Jannach2022}. Common methods like the paired Student’s t-test become more sensitive as the number of test users increases \citep{demvsar2006statistical}. Thus \textit{number of test users after data splitting should be sufficient} to provide a robust estimate and draw reliable conclusions.

\subsubsection{Splitting strategies for SRS in publications and RS frameworks}
\label{sec:papers_and_frame}
To assess the prevalence of data splitting practices, we analyzed recent SRS papers from conferences RecSys, SIGIR, and CIKM for 2022--2024. Out of 75 papers that conducted offline evaluation, 77.3\% used LOO-based splits, 16\% used GTS-based strategies, and only 6.7\% of works applied GTS in the next-item-prediction setting. This shows that splits, both preserving the global timeline and NIP task-oriented~\cite{harte2023leveraging,klenitskiy2024does,mezentsev2024scalable,frolov2024self}, remain rare within the RS community.

We also observe an absence of splits suitable for the SRS in popular RS frameworks. Table~\ref{tab:frameworks} shows that frameworks often offer LOO and GTS-based splits, but lack direct support for their combinations tailored for NIP with GTS. The absence of appropriate splits in widely adopted frameworks limits their usage in RS community.

\begin{table}[b]
\setlength{\abovecaptionskip}{3pt}
\caption{Splits available in popular RS frameworks} \label{tab:frameworks}
\resizebox{\columnwidth}{!}{%
\begin{tabular}{lcccccccc}
\toprule
\multirow{2}{*}{\textbf{Split Type}} & Cornac    & DaisyRec  & Elliot    & FuxiCTR   & Recomdrs.  & RecBole   & RecPack   & RePlay    \\
 &
  {\small \cite{salah2020}} &
  {\small \cite{sun2023}} &
  {\small \cite{anelli2021}} &
  {\small \cite{Jieming2021}} &
  {\small \cite{Graham2019}} &
  {\small \cite{Zhao2021}} &
  {\small \cite{Michiels2022}} &
  {\small \cite{Vasilev2024}} \\ \midrule
LOO-based                             & \cmark & \cmark & \cmark & \xmark & \cmark & \cmark & \cmark & \cmark \\
GTS-based                            & \xmark & \cmark & \cmark & \cmark & \cmark & \cmark & \cmark & \cmark \\
GTS + NIP                         & \xmark & \xmark & \xmark & \xmark & \xmark & \xmark & \cmark & \xmark \\ \hline
\end{tabular}%
}
\end{table}

\begin{figure}[t]
\centering
\setlength{\abovecaptionskip}{5pt}
\includegraphics[width=0.85\columnwidth]{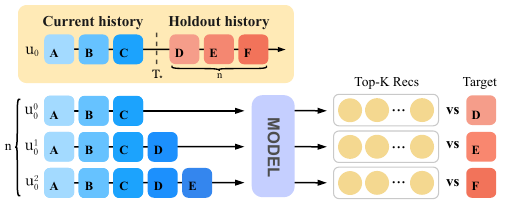}
    \caption{Successive evaluation scheme applied to one user with $n=3$ holdout interactions.}
\label{fig:succ}
\end{figure}

\subsubsection{Splitting strategies for SRS in detail}
In our work, we consider the leave-one-out split as the most popular in sequential recommendation research, and the global temporal split with the different target variants, either widespread or aligned with the next-item prediction task. We leave out of scope the other splitting strategies, based on the random, temporal user, or user split defined by \citet{meng2020exploring}, as they are less common for sequential recommendations, mismatch the task, and fail to preserve the global timeline.
\begin{itemize}[  leftmargin=*,
itemindent=6pt,label=\textbullet,nosep]
    \item \textbf{Leave-one-out (LOO)} holds out each user’s final interaction for testing and the second-to-last for validation (Fig.~\ref{fig:teaser}a). LOO supports local chronology, aligns with the NIP task, and in some cases maximizes the amount of training data and the number of test users. LOO is used in fundamental works~\cite{kang2018self, sun2019bert4rec}. However, LOO ignores a global timeline, allowing future data to leak into the training set~\citep{hidasi2023widespread,sun2023take}, and produces an unrealistically long test period, as target events span the entire dataset.
    
    \item \textbf{Global Temporal Split (GTS)} defines a global timepoint (cutoff) $T_{\text{test}}$ based on a quantile of interaction data or on a required number of test sequences, and assigns all interactions after $T_{\text{test}}$ to the holdout set (Fig.~\ref{fig:teaser}b).
    The GTS description for sequential recommendation is incomplete without specifying the target item selection strategy. Nevertheless, the general properties of GTS are as follows: GTS completely prevents data leakage and gives control over the test period duration. Compared to LOO, GTS typically produces fewer test users, and it often requires a dataset-specific choice of a global timepoint to balance the number of test users, test period duration, and amount of training data.
\end{itemize}
    
\paragraph{Global temporal split target options}
We propose to consider a wide range of target options for GTS illustrated in Figure~\ref{fig:teaser}c. The common approach
bundles \textit{all of a user’s holdout items} as a single ground-truth set \cite{gunawardana2012evaluating}, which we refer to as an \textbf{All} target.
The other considered targets are based on the following: \textit{the selected holdout item (or each of the holdout items) is treated as a separate target}. \textit{All interactions that occurred before the current target are available during inference}, regardless of whether they occur before or after the global split point, and \textit{all subsequent interactions are discarded}.
We consider the following target‐selection strategies aligned with the next-item prediction task:

\begin{itemize}[leftmargin=*,
itemindent=6pt,label=\textbullet,nosep]
  \item \textbf{Last}: the last holdout interaction is considered as a target. This target is an intuitive combination of LOO and GTS~\cite{klenitskiy2024does,harte2023leveraging, gusak2024rece}. This target emphasizes later post‐timepoint behavior.
  \item \textbf{First}: the first holdout interaction is considered as a target. Emphasizes early post‐timepoint behavior and may be skewed due to session-boundary artifacts (see Section~\ref{rq:deltas}). 
  \item \textbf{Successive (Sucv.)}: each holdout interaction is a separate target (Fig.~\ref{fig:succ})~\cite{frolov2024self,sun2023take}. As a result, for $n$ holdout interactions, we obtain $n$ separate metric evaluations. Then the metrics could be preliminarily averaged by user or directly used to estimate the mean metric value. Preliminary averaging by the user is more aligned with real-world usage, with each user being equally important, avoiding distribution bias towards active users who produce a significant number of interactions. 
  Successive target is the most aligned with the real-world SRS usage in the online scenario and represents an average performance after a global timepoint. Yet
  it is more complex to implement, and the average holdout sequence length linearly increases the inference computation compared to variants with one target item per sequence.
  \item \textbf{Random}: target interaction is randomly sampled from the holdout sequence. It provides a proxy for an average post-cutoff performance and can be considered an alternative to successive evaluation. However, due to its non-deterministic nature, ensuring reproducibility requires extra actions like averaging over multiple random seeds or publishing sampled targets.
\end{itemize}

\begin{figure}[t]
    \centering
\setlength{\abovecaptionskip}{4pt} 
\includegraphics[width=1\columnwidth]{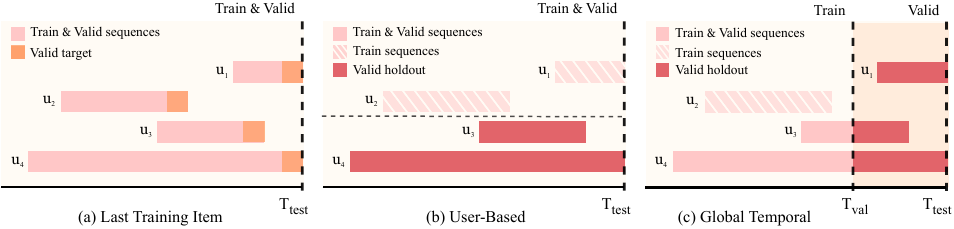}
    \caption{Validation split options for GTS (Fig.~\ref{fig:teaser}b): (a) each user's Last training item is a target, (b) User-based: interactions of $n$ random users are reserved for holdout, (c) Global temporal: interactions after $T_{\textnormal{val}}$ are reserved for holdout. Targets for holdout sequences are chosen according to Figure~\ref{fig:teaser}c.}
\label{fig:valids}
\end{figure}

\paragraph{Validation splitting strategies for GTS}
The most straightforward validation approach for GTS is global temporal validation. However, we also consider the other validation options that allow for the inclusion of more recent data before the test global timepoint into the training dataset. It provides better quality on the test subset without model retraining on the training and validation data combination. Although those options permit some temporal leakage into validation set, the final model ranking remains unaffected, since the ultimate evaluation metric is computed strictly on the GTS-based test set. We consider three validation options for GTS (Fig.~\ref{fig:valids}):
\begin{itemize}[leftmargin=*,
itemindent=6pt,label=\textbullet,nosep]
  \item \textbf{Global Temporal (GT)} sets a cutoff $T_{\mathrm{val}}$ before $T_{\mathrm{test}}$ and holds all interactions after $T_{\mathrm{val}}$ for validation. It prevents temporal leakage; matches test split, but shrinks the training set and drops recent user interactions.
  \item \textbf{Last Training Item (LTI)} holds each user’s final interaction before $T_{\mathrm{test}}$ as the  validation target. It covers all users and aligns with NIP but yields an unrealistically long validation period. 
  \item \textbf{User-Based (UB)} reserves the entire histories of a random subset of users for validation holdout. It preserves full histories for training users and limits training data reduction. Still, it yields an unrealistically long validation period and requires direct control of the number of users in training and validation.
\end{itemize}

For the UB and GT validation splitting strategies, there is a need for the target item selection, and options considered in \textit{Global Temporal Split Target Options}, especially the Last, Successive, and Random for UB, are also applicable for the validation subset.

\paragraph{Reporting of splitting strategy details} The necessity of careful description of splitting details is highlighted in multiple works on the topic~\cite{gunawardana2012evaluating,meng2020exploring}. In addition to details common to various splits and recommendation tasks (such as including the train-test split ratio and cold items filtering), we emphasize the need to report specific details for the sequential recommendation task. Those details depend on the chosen splitting strategy and may include the target item or item set selection, the input sequence building approach for selected targets, the presence of new sequences started after the global timepoint in holdout, and the target selection approach for these sequences. In Section \ref{sec:metrics}, we report details of the GTS splitting used in this study.

\section{Experiments}
\label{sec:expers}

We design experiments to answer the following research questions:
\begin{enumerate}[font={\bfseries}, label={RQ\arabic*}, wide, labelindent=0pt]
    \item What are the important properties of subsets obtained with different splitting strategies? 
    \item What is a distribution of time delta between consecutive user interactions, and how does it affect target item selection for GTS?
    \item How consistent are recommendation metrics for different splitting strategies in terms of correlation?
    \item How do different data splitting strategies influence the final model rankings?
    \item Which validation strategies are more appropriate for GTS?
    \item How does retraining the model on the combined training and validation data influence its final test performance?
    
\end{enumerate} \label{rqs}

The code for our experiments is available in the repository.\footnote{\url{https://github.com/monkey0head/time-to-split}\label{github}}

\setcounter{subsection}{-1} 
\subsection{Experimental Setup}
\label{sec:expsetup}

\subsubsection{Datasets}
We conduct our experiments on eight popular real-world datasets, mostly selected for their strong sequential structure, as highlighted in recent research~\cite{klenitskiy2024does}:
Amazon Reviews~\cite{mcauley2015imagebased} datasets (\textbf{Beauty}, \textbf{Sports}), MovieLens-1M (\textbf{ML-1M})
and MovieLens-20M (\textbf{ML-20M})~\cite{ml1}, 
BeerAdvocate (\textbf{BeerAdv})~\cite{mcauley2012beer}, 
\textbf{Diginetica}\footnote{\url{https://competitions.codalab.org/competitions/11161}\label{digi}}, \textbf{YooChoose}~\cite{yooooo}, and \textbf{Zvuk}~\cite{shevchenko2024variability}. 
To manage computational costs while ensuring sufficient data for analysis, we sample 2,000,000 users from the YooChoose dataset and 20,000 users from Zvuk.

Consistent with prior studies~\cite{tang2018personalized, klenitskiy2023turning}, we treat any review or rating as implicit feedback.
Additionally, following common practices \cite{sun2021areweevaluatingrigorouslypcore, sachdeva2020howusefularereviews, dacrema2019arewereallymakingmuchprogress}, we apply $p$-core filtering with $p$ equal $5$ to discard unpopular items and short user sequences.
Furthermore, we eliminate consecutive repeated items in user interaction histories~\cite{hidasi2023widespread}. 
Table~\ref{tab:dataset_stats} summarizes the final statistics of the datasets.

\begin{table}[t]
\setlength{\abovecaptionskip}{3pt}
\caption{Statistics of the datasets after preprocessing} \label{tab:dataset_stats}
\resizebox{0.9\columnwidth}{!}{%
    \centering
    \begin{tabular}{lrrrrrr}
    \toprule
        \textbf{Dataset} & \textbf{\#Interact.} & \textbf{\#Users} & \textbf{\#Items} & \textbf{Avg. Len.} & \textbf{Density (\%)} & \textbf{\#Days} \\ \midrule
        Beauty~\cite{mcauley2015imagebased} & 198 502 & 22 363 & 12 101 & 8.9 & 0.07 & 4 424 \\ 
        BeerAdv~\cite{mcauley2012beer} & 1 475 412 & 14 635 & 22 074 & 100.8 & 0.46 & 5 620 \\ 
        Diginetica\hyperref[digi]{\textsuperscript{\ref{digi}}} & 485 903 & 61 279 & 25 593 & 7.9 & 0.03 & 152 \\ 
        ML-1M~\cite{ml1} & 999 611 & 6 040 & 3 416 & 165.5 & 4.84 & 1 038 \\ 
        ML-20M~\cite{ml1}  & 19 984 024 & 138 493 & 18 345 & 144.3 & 0.79 & 7 385 \\ 
        Sports~\cite{mcauley2015imagebased} & 296 337 & 35 598 & 18 357 & 8.3 & 0.05 & 4 521 \\ 
        YooChoose~\cite{yooooo} & 2 792 229 & 335 203 & 20 758 & 8.3 & 0.04 & 181 \\ 
        Zvuk~\cite{shevchenko2024variability} & 8 087 953 & 19 267 & 150 206 & 419.8 & 0.28 & 91 \\ 
        \bottomrule
    \end{tabular}
    }
\end{table}

\subsubsection{Evaluation}
\label{sec:metrics}
For GTS, we use $q_{0.9}$ interaction quantile to conduct the main experiments. We filter out sequences of length one from all data subsets. For test and validation subsets, we use all sequence elements before the target item as an input in inference, regardless of their position relative to the global timepoint. For sequences started after the global timepoint, we excluded the first item from the targets to provide a model with at least one element of the sequence. For the same reason, we only use sequences that start before the global timepoint for the All target. We apply preliminary metric averaging within a sequence (user history) for the Successive target.
For GTS with GT and UB validation, we use the Last target as a reasonable and deterministic choice, allowing to reduce the training computational costs compared to the Successive target. For GTS with UB validation, we sample 1024 users. 

Recent studies highlighted the limitations of using sampled metrics for evaluating RS, as they can introduce biases and misrepresent model performance ~\cite{Dallmann_2021, rocio21, 50163}.
Following best practices, we use popular \emph{unsampled} top-K ranking metrics\footnote{We compute metrics using RePlay framework:
 \url{https://github.com/sb-ai-lab/RePlay}}: Normalized Discounted Cumulative Gain (NDCG@K), Mean Reciprocal Rank (MRR@K) and HitRate (HR@K), with K $= 5, 10, 20, 50, 100$.

In line with common practices, we also apply a \emph{filter seen}~\cite{filter_seen} step, removing items from the recommendation lists that users have already interacted with. This step is applied to all datasets except Zvuk, YooChoose, and Diginetica, as these datasets naturally contain repeated user-item interactions~\cite{klenitskiy2024does}.

\subsubsection{Models}
\label{sec:models}

We conduct our experiments using three popular sequential recommender system baselines: 
\textbf{SASRec$^+$}~\cite{klenitskiy2023turning}, an adaptation of the original PyTorch implementation\footnote{\url{https://github.com/pmixer/SASRec.pytorch}} that employs full cross-entropy loss (CE) over the entire item catalog to achieve state-of-the-art  performance~\cite{klenitskiy2023turning, petrov2023gsasrec};
\textbf{BERT4Rec}~\cite{sun2019bert4rec}, an efficient implementation\footnote{\url{https://github.com/antklen/sasrec-bert4rec-recsys23}\label{bertgru}} using the Transformers library~\cite{wolf2019huggingface};
and \textbf{GRU4Rec}~\cite{hidasi2015session} implementation\hyperref[bertgru]{\textsuperscript{\ref{bertgru}}} with full CE loss~\cite{klenitskiy2023turning}.

\subsubsection{Implementation Details} 
\label{sec:implementation_details}

In our experiments, we define wide ranges for each model’s hyperparameters~\cite{hidasi2023effect, klenitskiy2023turning,petrov2022systematic}. For both SASRec$^+$ and BERT4Rec, we vary the hidden sizes between 32 and 256, use between 1 and 3 self-attention blocks, and from 1 to 4 attention heads. We applied a masking probability of 0.2 for BERT4Rec. In the case of GRU4Rec, we explore hidden sizes from 16 up to 512 and vary the number of GRU layers from 1 to 4. We also employ dropout rates between 0.1 and 0.5 across all models.

For all models, we use a training batch size of 256 and set the maximum sequence length to 128. We train the models using the Adam optimizer with a learning rate of 10\small{$^{-3}$} \normalsize and set the maximum number of epochs to 300. 
During training, we monitor NDCG@10 on the validation set to control model convergence through the early stopping mechanism.
Specifically, we set the patience parameter to 10 epochs for SASRec$^+$ and GRU4Rec, while for BERT4Rec we use a patience of 20 to accommodate its slower convergence, observed in prior studies~\cite{klenitskiy2023turning,petrov2022systematic}.
All experiments are conducted on NVIDIA H100 GPUs with 80GB HBM3 memory.

\begin{table}[t!]
\setlength{\abovecaptionskip}{3pt}
\caption{Holdout statistics for different splits ($q_{0.9}$ for GTS)} \label{tab:valid_test_stats}
\resizebox{1\columnwidth}{!}{%
    \centering
\begin{tabular}{p{1cm}llrrrrrrrr}
\toprule
\multicolumn{1}{c}{\textbf{Set}} & \textbf{Split} & \textbf{Stats.\tiny{ $\downarrow$}} & 
{\small Beauty}    
  & {\small BeerAdv}   
  & {\small Diginetica} 
  & {\small ML-1M}     
  & {\small ML-20M}     
  & {\small Sports}     
  & {\small YooChoose} 
  & {\small Zvuk}     \\ \midrule 
\multirow{4}{*}{\parbox{1cm}{\centering Full\\Data}} & \multirow{4}{*}{$-$} & \#Days          & 4,424  & 5,620  & 152   & 1,038 & 7,385   & 4,521  & 181    & 91    
\\
                           &                      & Lifetime (\%)        & 12.4 & 11.56 & 0.01 & 9.14 & 2.66 & 12.0 & 0.01 & 43.47   \\
                           &                      & \#Users         & 22,363 & 14,635 & 61,279 & 6,040 & 138,493 & 35,598 & 335,203 & 19,267 \\
                           &                      & Seq. Len.       & 8.88  & 101   & 7.93  & 166  & 144    & 8.32  & 8.33   & 420   \\ 
                           
                           \midrule
\multicolumn{1}{c}{\multirow{9}{*}{Valid}}    & \multirow{2}{*}{LOO} & \#Days (\%)     & 84.0  & 66.9  & 100   & 100  & 94.7   & 69.4  & 100    & 100   \\
                           &                      & \#Users (\%)    & 100   & 100   & 100   & 100  & 100    & 100   & 100    & 100   \\ \cmidrule{2-11} 
                           & \multirow{3}{*}{GT}  & \#Days (\%)     & 1.38  & 2.76  & 6.58  & 2.41 & 10.9   & 1.50  & 8.29   & 7.69  \\
                           &                      & \#Users (\%)    & 28.0  & 35.4  & 9.55  & 17.3 & 11.9   & 27.2  & 8.87   & 41.7  \\
                           &                      & Holdout Len. & 2.84  & 25.6  & 7.47  & 86.0 & 109    & 2.73  & 8.45   & 90.7  \\ \cmidrule(l){2-11}
                           & \multirow{2}{*}{UB}  & \#Days (\%)     & 45.6  & 57.3  & 90.1  & 23.8 & 79.3   & 41.2  & 90.6   & 91.2  \\
                           &                      & \#Users (\%)    & 4.58  & 7.00  & 1.67  & 17.0 & 0.74   & 2.88  & 0.31   & 5.31  \\ \cmidrule(l){2-11}
                           & \multirow{2}{*}{LTI} & \#Days (\%)     & 82.4  & 63.7  & 94.1  & 23.8 & 79.6   & 66.1  & 90.6   & 91.2  \\
                           &                      & \#Users (\%)    & 96.0  & 94.0  & 90.0  & 99.5 & 90.2   & 96.1  & 90.3   & 95.6  \\ \midrule
\multicolumn{1}{c}{\multirow{5}{*}{Test}}      & \multirow{2}{*}{LOO} & \#Days (\%)     & 84.0  & 66.9  & 100   & 100  & 94.5   & 68.1  & 100    & 100   \\
                           &                      & \#Users (\%)    & 100   & 100   & 100   & 100  & 100    & 100   & 100    & 100   \\ \cmidrule{2-11} 
                           & \multirow{3}{*}{GTS} & \#Days (\%)     & 1.60  & 3.26  & 5.92  & 76.1 & 14.9   & 1.95  & 9.39   & 8.79  \\
                           &                      & \#Users (\%)    & 27.3  & 35.0  & 10.4  & 20.0 & 13.4   & 28.7  & 9.74   & 43.8  \\
                           &                      & Holdout Len. & 3.25  & 28.8  & 7.66  & 82.7 & 108    & 2.89  & 8.55   & 95.9  \\ \bottomrule
\end{tabular}%
}
\end{table}

\subsection{Split statistics and properties (RQ1)}
\label{rq:statistics}

Different splitting strategies generate different training, test, and validation subsets, which could vary significantly. In this section, we analyze important subsets' properties and the influence of the splitting strategy on them.

\subsubsection{Amount of training data left} 
GTS offers \textit{direct control of the amount of training data}; thus, for the $0.9$-quantile, 90\% of data points are left for training and validation.
It seems intuitive that the LOO split should leave more data for training, as it does not preserve the global timeline and holds only two items of each sequence. However, for the datasets with short sequences (Beauty, Sport, Diginetica, YooChoose), LOO leaves about 75\% of interactions for training, less than any GTS variant. 
It should be noted that \textit{splitting also affects average sequence length}. Thus, for the datasets with a long user lifetime, calculated as a median period of user activity divided by the dataset time period (Table \ref{tab:valid_test_stats}, Lifetime (\%)), GTS shortens training sequences by roughly 20\% in Beauty, Sports and Zvuk, while lengths in Diginetica and YooChoose remain nearly unchanged. 

\subsubsection{Trade-off between the number of test users, the volume of training data, and the duration of the test period}

Table~\ref{tab:valid_test_stats} shows that LOO includes 100\% of users in the test set, whereas GTS at $q_{0.9}$ covers only 10\%–44\%. Thus, it could be easier to obtain statistically significant results with LOO, but those results are obtained in an unrealistic setup, not preserving the global timeline. Table~\ref{tab:test_quant} compares GTS across quantiles $\{0.8,0.9,0.95,0.975\}$, revealing up to a 4 times decrease in test users at $q_{0.975}$. Lower quantiles raise user counts but extend the test subset duration, making it unrealistically long. Holdout period for GTS at $q_{0.9}$ spans 2\%–15\% of the timeline (8–1,100 days) and rises to 100\% under LOO (7,385 days for ML-20M). For Zvuk and Diginetica, GTS at $q_{0.9}$ yields nearly a week-long test with thousands of users, matching real usage time period and users sufficiency requirements. 

The holdout length per user is also affected by the split. For LOO, it is always equal to one, while GTS holdout length varies with user lifetime and quantile. Table~\ref{tab:test_quant} reports holdouts exceeding 100 items in Zvuk and ML-20M, which significantly increases inference cost for successive evaluation. 
However, even for the higher quantiles or specific datasets like ML-1M, we observe an unrealistically long test period in some cases, combined with a lack of users. We recommend using datasets and GTS quantiles that balance user count, test duration, and training data amount.

\subsubsection{Influence of validation type on validation and training subsets properties} 
\label{sec:stats_validation_types}

 Table~\ref{tab:valid_test_stats} shows that test and validation subsets for LOO and for GTS with global temporal validation yield aligned holdout durations and user shares, thus those strategies could provide better validation and test metrics compliance. In contrast, the Last Training Item and User-Based validation yield subsets of a long validation period, less aligned with GTS test set statistics. Since UB reserves a user subset, its size should be additionally controlled to balance the training data amount and the number of validation users.

\begin{table}[t]
\setlength{\abovecaptionskip}{4pt}
\setlength{\abovecaptionskip}{3pt}
\caption{Test subset statistics for GTS for different quantiles} \label{tab:test_quant}
\resizebox{1\columnwidth}{!}{%
    \centering
\begin{tabular}{l|lrrrr|lrrrr|rrrrr}
\toprule
\multirow{2}{*}{\textbf{Dataset}} &
  \textbf{Len.} &
  \multicolumn{4}{c|}{\textbf{Holdout Len.}} &
  \multicolumn{5}{c|}{\textbf{\#Users (K)}} &
  \multicolumn{5}{c}{\textbf{\#Days}} \\ \cline{2-16} 
 &
  \textit{Full} &
  $q_{0.8}$ &
  $q_{0.9}$ &
  $q_{0.95}$ &
  $q_{0.975}$ &
  \textit{Full} &
  $q_{0.8}$ &
  $q_{0.9}$ &
  $q_{0.95}$ &
  $q_{0.975}$ &
  \textit{Full} &
  $q_{0.8}$ &
  $q_{0.9}$ &
  $q_{0.95}$ &
  $q_{0.975}$ \\ 
\midrule
Beauty     &  8.88  & 3.88  & 3.25  & 2.76  & 2.45  & 22.4 & 10.2 & 6.11  & 3.52  & 1.91  & 4,424 & 138 & 71 & 35 & 19 \\
BeerAdv    & 101    & 42.5  & 28.8  & 18.7  & 12.0  & 14.6 & 6.94  & 5.12  & 3.94  & 3.07  & 5,620 & 354 & 183 & 94 & 48 \\
Diginetica &  7.93  & 7.68  & 7.66  & 7.38  & 6.55  & 61.3 & 12.7 & 6.35  & 3.29  & 1.86  &   152 &  20 & 9  & 4  & 2  \\
ML-1M      & 166    & 112   & 82.7  & 61.5  & 45.6  &  6.04 & 1.78  & 1.21  & 0.81  & 0.55  & 1,038 & 818 & 790 & 617 & 400 \\
ML-20M     & 144    & 126   & 108   & 92.8  & 86.9  &139 & 31.7 & 18.6 & 10.8 & 5.75  & 7,385 &1,994&1,100& 569 & 201 \\
Sports     &  8.32  & 3.52  & 2.89  & 2.61  & 2.60  & 35.6 & 16.7 & 10.2 & 5.63  & 2.79  & 4,521 & 163 & 88 & 43 & 22 \\
YooChoose  &  8.33  & 8.49  & 8.55  & 8.57  & 8.79  &335 & 65.8 & 32.7 & 16.3 & 7.94  &   181 &  34 & 17 & 10 & 5  \\
Zvuk       & 420    & 150   & 95.9  & 61.6  & 42.8  & 19.3 & 10.8 & 8.43  & 6.57  & 4.73  &    91 &  16 & 8  & 4  & 2  \\ 
\bottomrule
\end{tabular}%
}
\end{table}

\subsection{Time gaps for different targets in GTS (RQ2)}
\label{rq:deltas}

\begin{table}[t]
\setlength{\abovecaptionskip}{3pt}
\caption{Median delta $\delta$ (in seconds) between each target interaction and the previous one: for different (a) validation types on the validation, and (b) target options on the test set}
\label{tab:test_and_val_delta}
\resizebox{\columnwidth}{!}{%
\begin{tabular}{llrrrrrrrr}
\toprule
\textbf{Set} & \textbf{Setup\tiny{ $\downarrow$}}   & {\small Beauty}    
  & {\small BeerAdv}   
  & {\small Diginetica} 
  & {\small ML-1M}     
  & {\small ML-20M}     
  & {\small Sports}     
  & {\small YooChoose} 
  & {\small Zvuk}     \\ \midrule
Full Data & $-$ & 345,600 & 73,182 & 58 & 0 & 11 & 172,800 & 59 & 14  \\ 
\midrule
\multirow{4}{*}{(a) Valid} & LOO   & 172,800   & 360,900 & 63 & 18 & 17 & 86,400    & 59 & 98  \\
                       & GT Last    & 1,036,800 & 446,371 & 71 & 27 & 41 & 1,209,600 & 67 & 84  \\
                       & UB    &   604,800 & 691,188 & 70 & 15 & 19 &   518,400 & 65 & 78  \\
                       & LTI   &   604,800 & 690,794 & 70 & 15 & 21 &   518,400 & 65 & 68  \\ \midrule
\multirow{5}{*}{(b) Test}  & LOO   &   604,800 & 737,140 & 70 & 17 & 20 &   518,400 & 65 & 73  \\
                       & Last  & 1,382,400 & 508,452 & 70 & 67 & 29 & 1,296,000 & 68 & 91  \\
                       & First & 8,640,000 & 4,921,729 & 186 & 7,153,214 & 21,145,894 & 11,577,600 & 259 & 346,010 \\
                       & Rand. & 3,628,800 &   439,805 & 65 & 35 & 15 &   4,752,000 & 62 & 120 \\
                       & Succ. &   172,800 &    75,916 & 58 & 22 & 14 &      86,400 & 60 & 67  \\ \bottomrule
\end{tabular}%
}
\end{table}

\subsubsection{Temporal distribution of user activity inside sequence}
Experimenting with real-world and some academic datasets, we observed that the First interaction after the global timepoint as the GTS target yields lower metrics than for subsequent items. We hypothesized that the global timepoint often hits a period of user's inactivity, an inter-session period, and thus the First item becomes the beginning of the next user session. In our work, we do not explicitly identify sessions, as it is often a matter of professional judgment (heuristic) \citep{Jannach2022}, but in Table~\ref{tab:test_and_val_delta}, we report median time gaps between all consecutive interactions in datasets (indicated as Full Data) and each target and its previous event. The gap for the First target is much larger than for other targets, and the time gap across the dataset, which makes this target biased.  Thus, we do not recommend using the First item after the global timepoint as a target.

\label{rq:test_vs_test_correlations}

\subsubsection{Time gap patterns across targets}
\begin{figure}[b!]
    \centering
    \setlength{\abovecaptionskip}{-4pt} 
    \setlength{\belowcaptionskip}{0pt} 
 \includegraphics[width=0.97\columnwidth]{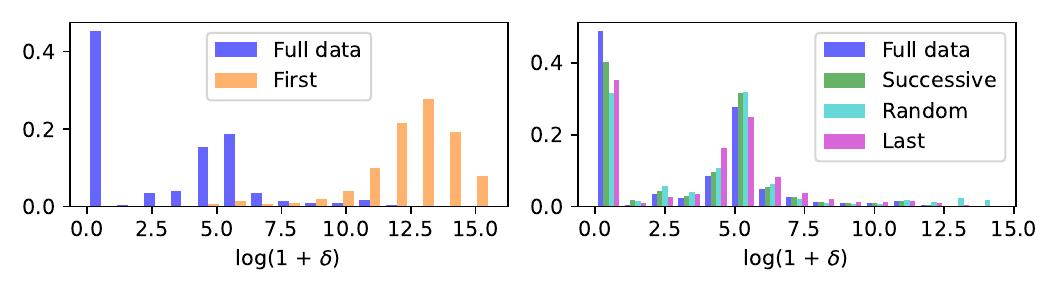}
        \caption{Distribution of time gaps $\delta$ between all interactions (Full data), and between target interaction and previous one for the First (left) and all others (right) target options for GTS on Zvuk dataset.} 
    \label{fig:time_delta}
\end{figure}

Table~\ref{tab:test_and_val_delta} shows that some review datasets (Beauty, Sports, BeerAdvocate) have day‐level median between-interactions time gaps, while the other datasets have second- or minute-level gaps that remain consistent across validation setups and targets except the First. On Beauty and Sports, short holdouts could lead to selection of the same (first) item as First, Last, and Random target, inflating the gap for Last and Random targets. As shown in Figure~\ref{fig:time_delta}, which plots the log‐scaled gap densities for Zvuk, the First target distribution is shifted right, while other targets match the overall pattern. This confirms that, except for the First interaction, all targets after the global timepoint are appropriate in terms of temporal intervals between interactions.

\subsection{Consistency between different splits (RQ3)}

\begin{figure}[ht!]
    \centering
    \setlength{\abovecaptionskip}{0pt} 
    \includegraphics[width=1.0\linewidth]{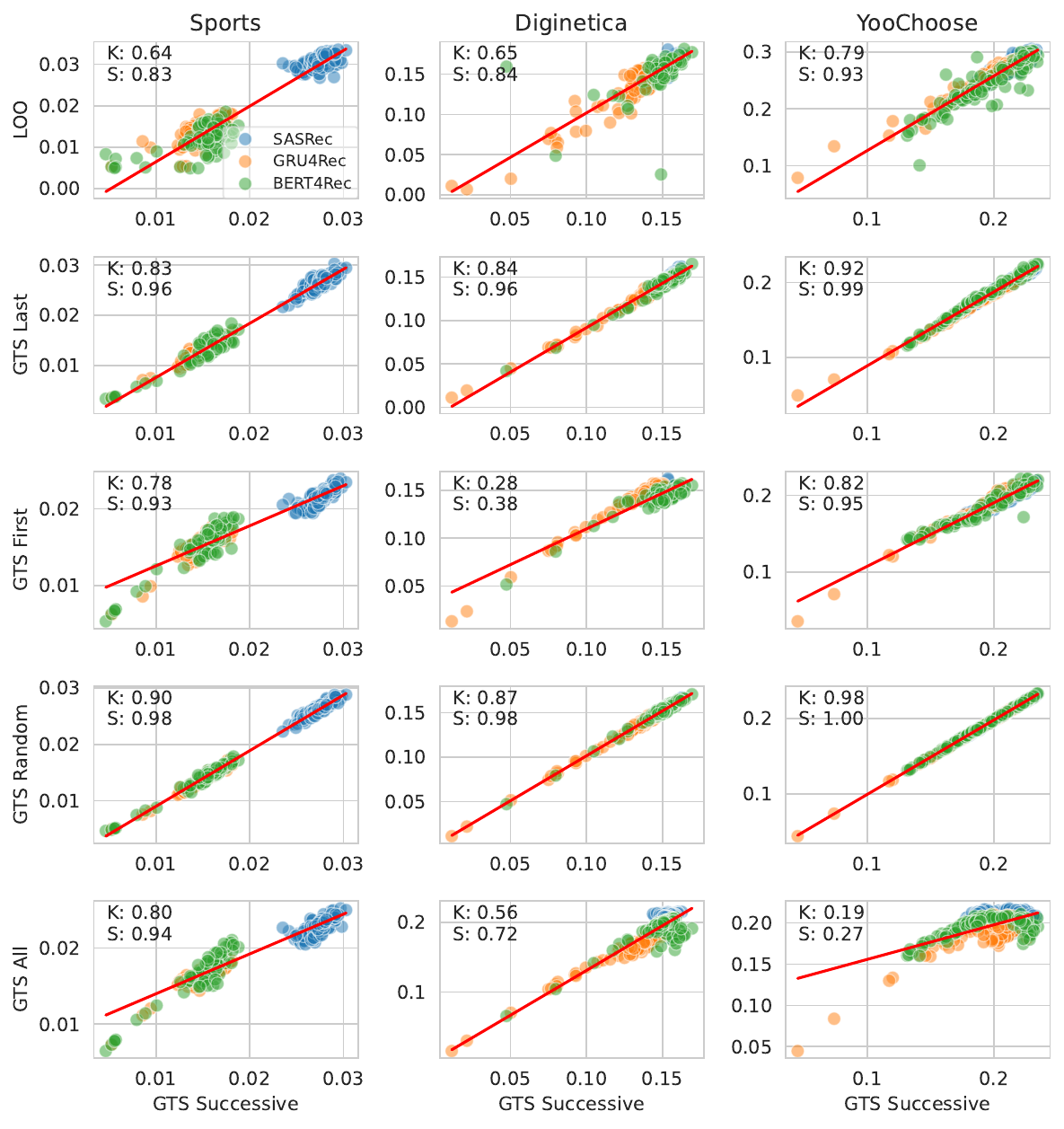}
        \caption{Scatterplots for NDCG@10 between GTS Sucv. target and other options. \textit{K} and \textit{S} denote Kendall and Spearman.} 
    \label{fig:test_vs_test_scatterplot}
\end{figure}

\begin{figure*}[ht!]
\setlength{\abovecaptionskip}{0pt} 
    \centering
    \includegraphics[width=1.0\textwidth]{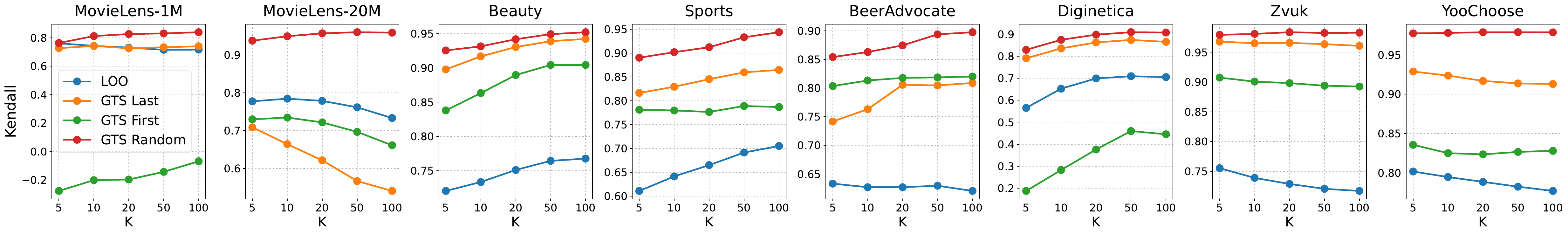}
        \caption{Kendall correlation between test NDCG@K for GTS with Successive target and other options.} 
    \label{fig:test_vs_test_corr}
\end{figure*}

Absolute metric values can vary significantly across different splitting strategies, making direct comparison of the results impossible. However, if the relative ranking of models is preserved across splits, then conclusions about their comparative performance remain consistent. To analyze the agreement between pairs of splits, we compute the correlation between metrics obtained on these splits across different models and hyperparameter settings. We treat the GTS with Successive target as the most realistic and closest to production use, and compare all other splits against it, suggesting that an appropriate split for next-item prediction should exhibit high correlation with this reference.

We train the models with a wide range of hyperparameters defined in Section~\ref{sec:implementation_details}, resulting in 108 configurations for SASRec$^+$ and BERT4Rec, and 104 for GRU4Rec. Using multiple hyperparameter settings for each model allows us to generate a large number of evaluation points, leading to more statistically robust conclusions. For a comprehensive analysis, we consider multiple evaluation metrics (HR, MRR, NDCG) at different values of K. To assess the agreement between the metrics obtained from different splits, we use Kendall and Spearman rank correlation coefficients.

\subsubsection{Visual scatterplot analysis.}

The first step is a visual analysis of scatter plots for different pairs of splits. Figure~\ref{fig:test_vs_test_scatterplot} shows pairwise comparisons between GTS Successive and other splits for NDCG@10 on several datasets. The highest correlation is observed for GTS Random. GTS Last also shows a high, though slightly lower, correlation. In contrast, LOO, GTS First, and GTS All exhibit much greater dispersion across all datasets.

\subsubsection{Correlation at different values of K}

Figure~\ref{fig:test_vs_test_corr} shows Kendall correlation across different values of K for NDCG on all datasets. To improve clarity, we omit GTS All from the plots due to its much lower correlation.

The relative order between splits remains fairly stable across different values of K. GTS Random shows the highest agreement with GTS Successive split, which is expected given their structural similarity. GTS Last typically follows with slightly lower, but still strong, correlation. In contrast, LOO and GTS First consistently show lower correlation, with significant drops on some datasets.

An exception is observed on MovieLens datasets: GTS Last is not better than LOO on ML-1M, and performs noticeably worse on ML-20M. Still, GTS Random outperforms LOO on both, suggesting that these datasets may have biased distributions for the last interaction in user histories.

\subsubsection{Aggregated results}
\label{rq4:aggregated_results}

\begin{table}[t!]
\setlength{\abovecaptionskip}{4pt}
\caption{Mean (across datasets) correlations between test GTS Successive target and other options for different metrics. Best values are in bold, second best are underlined.}
\label{tab:test_vs_test_corr}
\resizebox{0.9\columnwidth}{!}{%
    \centering
    \begin{tabular}{lcccccc}
    \hline
    \multirow{2}{*}{\textbf{Test Split}} & \multicolumn{3}{c}{\textbf{Kendall}} & \multicolumn{3}{c}{\textbf{Spearman}} \\ 
    \cmidrule(lr){2-4} \cmidrule(lr){5-7}
    & HR@10 & MRR@10 & NDCG@10 & HR@10 & MRR@10 & NDCG@10 \\ 
    \hline
    LOO       & 0.71      & 0.70       & 0.71        & 0.87       & 0.86        & 0.87        \\
    GTS Last   & \underline{0.83} & \underline{0.82}  & \underline{0.83}   & \underline{0.93}  & \underline{0.94}   & \underline{0.94}   \\
    GTS First  & 0.70      & 0.60       & 0.62        & 0.82       & 0.70        & 0.72        \\
    GTS Random & \textbf{0.91} & \textbf{0.90} & \textbf{0.91} & \textbf{0.98} & \textbf{0.98} & \textbf{0.98} \\
    GTS All    & 0.57      & 0.37       & 0.43        & 0.68       & 0.46        & 0.53        \\
    \hline
    \end{tabular}
}
\end{table}

Table~\ref{tab:test_vs_test_corr} presents Kendall and Spearman coefficients for HR@10, MRR@10, and NDCG@10 averaged across all datasets. The different metrics and correlation types show consistent trends. GTS Random achieves the highest average correlation, with GTS Last slightly behind. LOO performs significantly worse, then goes GTS First, while GTS All shows the lowest correlation, highlighting the task mismatch for this target.

To summarize, we conclude that GTS Last and GTS Random are suitable options for the evaluation of sequential recommendation models. Both can serve as reasonable alternatives to the more computationally expensive GTS Successive strategy. GTS Random shows the highest correlation, but it is non-deterministic, which can lead to reproducibility issues unless the exact splits are stored. Alternatively, GTS Random can be run multiple times with different seeds to obtain more stable average results and an estimate of metrics variability. The very low correlation of GTS All confirms that it significantly deviates from the next-item prediction objective. GTS First also proves to be a less correlated target, consistent with the analysis presented in Section~\ref{rq:deltas}. 
Finally, the experimental results support the assumption of the limited alignment of the commonly used LOO split with a close-to-reality evaluation protocol.

\subsection{Model rankings across different splits  (RQ4)}
\label{rq:}

Accurate final model ranking is crucial for both reliable research outcomes and practical deployment decisions.
Although our correlation analysis (Section~\ref{rq:test_vs_test_correlations}) shows that model rankings vary depending on the data split, those shifts may be primarily related to lower-ranked models rather than the best-performing configurations.
In this section, we analyze the consistency of best model rankings across different splitting strategies~\cite{meng2020exploring}.

\begin{figure}[t!]
    \centering
\setlength{\abovecaptionskip}{4pt} 
\includegraphics[width=0.95\columnwidth]{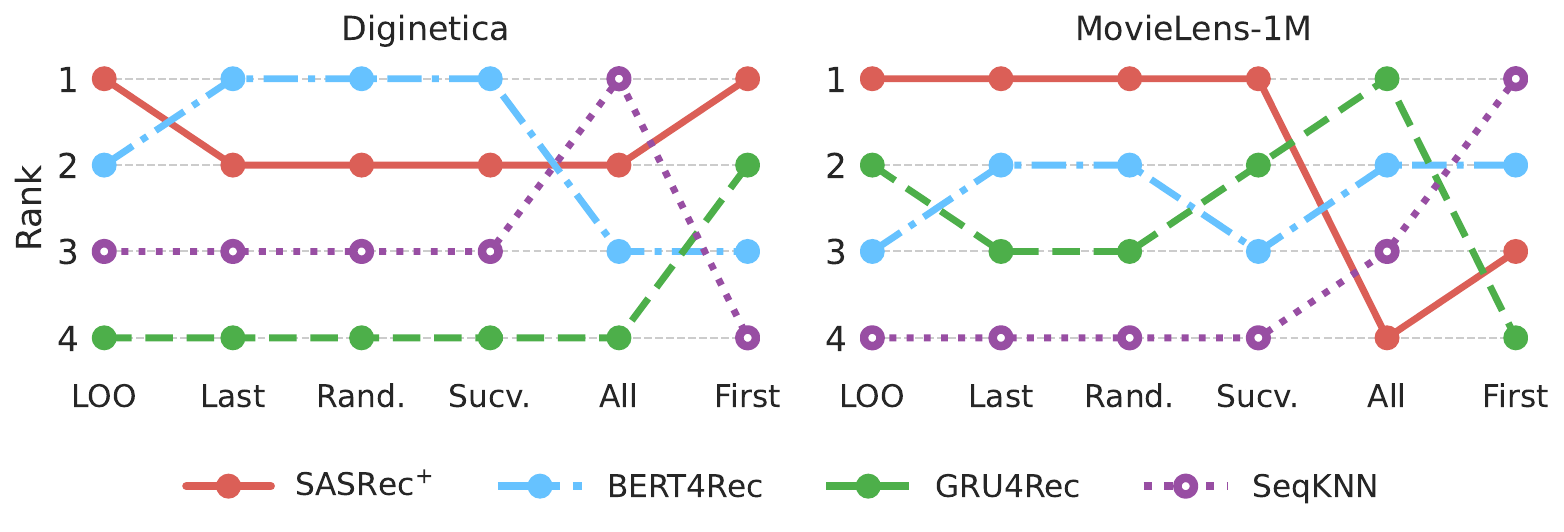}
    \caption{Model rankings based on test NDCG@10 for LOO split, and GTS split with global temporal validation.}
\label{fig:ranks}
\end{figure}

For each split and target type, we sort models by their best test performance and then track how their positions change across splits and GTS targets.
For this analysis, we also include \emph{sequential item-based kNN (SeqKNN)}~\cite{ludewig2018evaluation, seqknn} (a non-neural baseline) to better illustrate shifts in rankings.
Figure~\ref{fig:ranks} shows rankings by best test NDCG@10 under the LOO split, and the GTS split with GT validation and different test targets. 
We observe that model orderings \textit{regularly shift} when splits and GTS targets change, revealing unstable rankings.
For example, SASRec$^+$, which ranks first under the Successive target on ML-1M, falls to last place when evaluated under the All target option.
Such inconsistency holds across most datasets, except for Amazon Beauty and Sports, where rankings remain stable across evaluation targets.
Overall, SASRec$^+$ demonstrates the strongest performance on average, while GRU4Rec and SeqKNN frequently occupy the lowest positions.
Among different evaluation targets, Last, Random, and Successive yield comparable rankings, closely matched by LOO. In contrast, First and All produce noticeably different model orders.
Similar patterns hold across different test target options in alternative GTS validation setups (UB, LTI) and for different metrics.
In summary, our findings confirm that \emph{the choice of split and target may introduce significant ranking instability}. Careful selection of evaluation protocols is therefore essential for fair and reproducible comparisons.

\subsection{Validation strategies for GTS (RQ5)}
\label{rq:test_vs_validation_corr}
\setlength{\abovecaptionskip}{0pt} 
\begin{figure*}[ht!]
    \centering
    \includegraphics[width=1.0\textwidth]{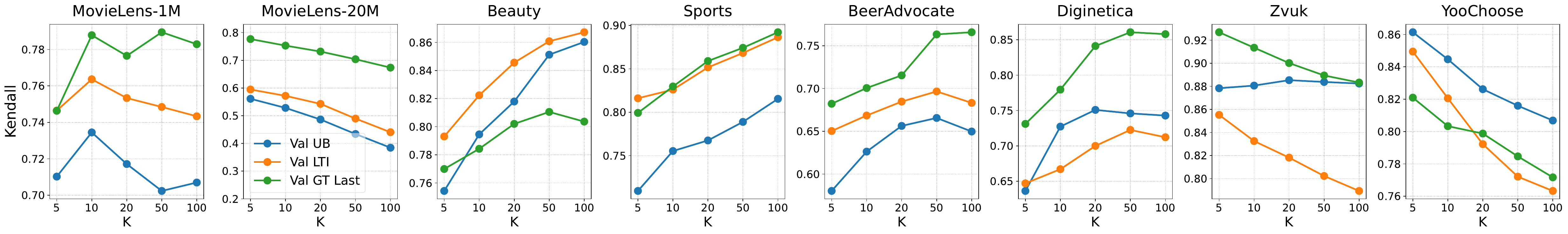}
        \caption{Kendall correlation between test and validation NDCG@K for GTS Last split with different validation strategies.} 
    \label{fig:test_vs_validation_corr}
\end{figure*}

\begin{table}[]
\setlength{\abovecaptionskip}{4pt}
\caption{Mean (across datasets) correlations between test and validation metrics for GTS with (a) \textnormal{\emph{Last}} and (b) \textnormal{\emph{Successive}} test targets and different validation types. Best values are in bold, second best are underlined.}
\label{tab:validation_corr_successive}
\resizebox{1\columnwidth}{!}{%
    \centering
    \begin{tabular}{ll|ccc|ccc}
    \toprule
    \multicolumn{2}{l|}{\textbf{Correlation}} & \multicolumn{3}{c|}{\textbf{Kendall}} & \multicolumn{3}{c}{\textbf{Spearman}} \\ \hline
    \textbf{Target} & \textbf{Valid. Type\tiny{$\downarrow$}} & {\small HR@10} & {\small MRR@10} & {\small NDCG@10} & {\small HR@10} & {\small MRR@10} & {\small NDCG@10} \\ 
    \midrule
    \multirow{7}{*}{(a) Test Last}   & UB            & 0.72 & 0.72 & 0.74 & 0.87 & 0.88 & 0.89 \\
                                     & LTI           & 0.73 & 0.75 & 0.75 & 0.88 & 0.90 & 0.90 \\
                                     & GT Last       & \textbf{0.78} & \textbf{0.79} & \textbf{0.79} & \textbf{0.93} & \textbf{0.93} & \textbf{0.93} \\
                                     & GT First      & 0.61 & 0.54 & 0.57 & 0.77 & 0.69 & 0.73 \\
                                     & GT Rand.     & 0.75 & 0.75 & 0.76 & 0.90 & 0.91 & \underline{0.92} \\
                                     & GT Sucv. & \underline{0.76} & \underline{0.77} & \underline{0.77} & \underline{0.91} & \underline{0.92} & \underline{0.92} \\
                                     & GT All        & 0.46 & 0.37 & 0.43 & 0.59 & 0.50 & 0.56 \\
    \midrule
    \multirow{7}{*}{(b) Test Sucv.}  & UB            & 0.78 & 0.78 & 0.80 & 0.93 & 0.92 & \underline{0.94} \\
                                     & LTI           & 0.80 & \textbf{0.83} & \underline{0.82} & \underline{0.94} & \textbf{0.95} & \textbf{0.95} \\
                                     & GT Last       & \underline{0.81} & \underline{0.81} & \underline{0.82} & \underline{0.94} & \underline{0.94} & \textbf{0.95} \\
                                     & GT First      & 0.64 & 0.56 & 0.59 & 0.80 & 0.72 & 0.75 \\
                                     & GT Rand.     & 0.80 & \underline{0.81} & 0.81 & \underline{0.94} & \underline{0.94} & \underline{0.94} \\
                                     & GT Sucv. & \textbf{0.83} & \textbf{0.83} & \textbf{0.83} & \textbf{0.95} & \textbf{0.95} & \textbf{0.95} \\
                                     & GT All        & 0.48 & 0.37 & 0.44 & 0.60 & 0.49 & 0.56 \\
    \bottomrule
    \end{tabular}%
}
\end{table}

While the validation choice for the LOO split is straightforward, the GTS split allows for multiple validation strategies, as described in Section~\ref{sec:strategies}.
To compare these strategies, we follow the same approach as in Section~\ref{rq:test_vs_test_correlations}, but now we examine how well the validation metrics align with the corresponding test metrics under a given split. 
If the correlation between validation and test performance  is low, the validation method is unreliable, as it may lead to selecting a suboptimal model.

\subsubsection{Correlation at different values of K}

Figure~\ref{fig:test_vs_validation_corr} shows the Kendall correlation between test and validation NDCG at various K across all datasets. The results are reported for GTS Last test split with three validation types: UB (user-based), LTI (last train item), and GT Last (global temporal with the last item as the target). For 5 out of 8 datasets, GT Last validation has consistently higher correlations with test metrics. On the Sports dataset, LTI validation performs on par with GT Last; on Youchoose, UB validation performs similarly to GT Last. On the Beauty dataset, both LTI and UB options show higher correlations than GT Last.

\subsubsection{Aggregated results}

We also compute the average correlation across all datasets for HR@10, MRR@10, and NDCG@10 similarly to Section~\ref{rq4:aggregated_results}. Table~\ref{tab:validation_corr_successive} reports the average Kendall and Spearman correlations for two test split types (GTS Last and GTS Successive) and all considered validation strategies. As expected, GTS test splits align best with the appropriate global temporal validation types (GT Last and GT Successive). The All and First validation targets perform the worst, for the same reasons discussed in the analysis of test splits. It is worth noting that LTI and UB validation lag only slightly behind global temporal options. However, they share similar drawbacks with the LOO split, as described in Section~\ref{sec:stats_validation_types}, and should therefore be used with caution. Another observation is that for the GTS Successive test split, the GT Last validation performs nearly as well as GT Successive, suggesting that the simpler GT Last validation can be used without a significant loss in reliability.

In summary, the experimental results suggest that under global temporal splitting, the most reliable validation strategy is the corresponding global temporal validation. However, the GT validation comes with a limitation: the most recent data is not used for training, as it is reserved for validation. As a result, the test performance can become lower compared to other validation strategies. 
The following section analyses the impact of retraining.

\subsection{Model retraining on combined data (RQ6)}
\label{rq:}
Retraining the optimal model on combined training and validation data adds complexity to the pipeline, but is essential to deliver peak performance at industrial deployment.
However, academic studies often omit retraining or leave it unreported, raising questions about the consistency of the results with real-world scenarios.

We select the best model on validation and compare the corresponding test metrics with and without retraining.
Table~\ref{tab:retrain} shows results for \emph{Successive} and \emph{Last} target options; other targets follow similar trends.
Without retraining, UB consistently outperforms GT and LTI on the test set.
We then examine the relative change in test metrics after retraining. The GT setup stably shows the largest improvement (e.g. 0.022 $\rightarrow$ 0.040: +81.8\% on Beauty for \textit{Successive}), as the retrained model captures shifts in user preferences over a long validation period, and benefits from additional training data. 
For UB, on average, retraining yields a modest improvement, while LTI and LOO experience more frequent performance drops. 
After retraining, \textit{GT and UB achieve similar absolute test scores} (e.g. 0.160 vs. 0.158 on Diginetica for \textit{Successive}), whereas LTI regularly remains lower.
We also perform correlation analysis, which, for both GT and UB, shows high Kendall's $\tau$ (0.6--0.9) and Spearman (0.7--1.0) between \emph{Test} and \emph{Test R.} metrics, indicating \textit{strong consistency for base and retrained setups} in selecting the same optimal models.
In contrast, LTI generally shows lower values (0.4--0.6 and 0.5--0.7).

Thus, when using GTS with GT or UB validation, retraining on combined training and validation data is important for achieving optimal deployment performance.
For academic comparisons, retraining is still recommended, but it does not substantially alter the relative ranking of models.

\begin{table}[t]
\setlength{\abovecaptionskip}{4pt}
\caption{Validation and test NDCG@10 of SASRec$^+$ at optimal validation configuration for different splits.
\textnormal{\emph{Test R.}} denotes setup with retraining on combined training and validation data.
\textnormal{\emph{LTI}} and \textnormal{\emph{UB}} in this study use only \textnormal{\emph{Last}} validation target. 
}
\label{tab:retrain}
\resizebox{0.93\columnwidth}{!}{%
\begin{tabular}{ll|cccr|cccr}
\toprule
\multicolumn{2}{l|}{\textbf{Dataset}}       & \multicolumn{4}{c|}{Diginetica}        & \multicolumn{4}{c}{Amazon Beauty}      \\ 
\hline
\textbf{Split} & \textbf{Target\tiny{ $\downarrow$}} & \textbf{Valid} & \textbf{Test} & \textbf{Test R.} & \textbf{$\Delta$ Test} & \textbf{Valid} & \textbf{Test} & \textbf{Test R.} & \textbf{$\Delta$ Test} \\ 
\midrule 
\multirow{2}{*}{GT} 
  & Last & 0.154 & 0.154 & 0.161 & 4.55\%  & 0.046 & 0.024 & 0.037 & 54.2\% \\
  & Sucv. & 0.154 & 0.149 & 0.160 & 7.38\%  & 0.044 & 0.022 & 0.040 & 81.8\% \\ 
\midrule
\multirow{2}{*}{UB}  
  & Last & 0.180 & 0.152 & 0.155 & 1.97\%   & 0.074 & 0.036 & 0.037 & 2.78\% \\
  & Sucv. &   –   & 0.159 & 0.158 & -0.63\% &   –   & 0.040 & 0.040 & 0.00\% \\ 
\midrule
\multirow{2}{*}{LTI} 
  & Last & 0.187 & 0.135 & 0.126 & -6.67\%  & 0.067 & 0.031 & 0.036 & 16.1\% \\
  & Sucv. &   –   & 0.147 & 0.129 & -12.2\%  &   –   & 0.036 & 0.039 & 8.33\% \\ 
\midrule 
LOO                  
  & Last & 0.179 & 0.181 & 0.157 & -13.3\% & 0.073 & 0.059 & 0.065 & 10.2\% \\ 
\bottomrule
\end{tabular}%
}
\end{table}

\section{Conclusion}
\label{sec:conclusion}

We systematically compared leave-one-out and global temporal splitting strategies with various validation types and evaluation targets for sequential recommendations. Our experiments show that the common leave-one-out split, besides allowing for the emergence of temporal leakage and criticism raised in previous studies, demonstrates lower correlation with real-world evaluation scenarios and can distort model rankings. 
We also proved that the GTS All target option suffers from a task mismatch with standard next-item prediction, and that GTS First exhibits lower correlation with more realistic evaluation strategies due to significant shifts in time-gap distributions between interactions.
In contrast, GTS with Last or Random target yields strong agreement with the more comprehensive but close-to-reality Successive evaluation scheme.
To summarize, we conclude that global temporal split with Last, Random, and Successive targets are appropriate options for the evaluation of sequential recommendation models, with the Last and Random being reasonable alternatives to the more computationally expensive Successive strategy. 

We further demonstrated that using a matching global temporal validation split produces reliable model selection, and that retraining on the combined training and validation data boosts final test performance
for the reasonable validation options,
compared to results for unretrained models.

\begin{acks}

We thank Fedor Dergachev for providing auxiliary code.
\end{acks}

\bibliographystyle{ACM-Reference-Format}
\balance
\bibliography{sections/bibliography}

\end{document}